# Multi-Band Superconductivity and the Steep Band/Flat Band Scenario

**Annette Bussmann-Holder** [1,*], **Hugo Keller** [2], **Arndt Simon** [1] **and Antonio Bianconi** [3,4,5]

[1] Max-Planck-Institute for Solid State Research, Heisenbergstr. 1, D-70569 Stuttgart, Germany; A.Simon@fkf.mpg.de

[2] Physik-Institute of the University of Zürich, University of Zürich, Winterthurerstr. 190, CH-8057 Zürich, Switzerland; keller@physik.uzh.ch

[3] RICMASS, Rome International Center for Materials Science Superstripes, Via dei Sabelli 119A, I-00185 Rome, Italy; antonio.bianconi@ricmass.eu

[4] Institute of Crystallography, Consiglio Nazionale delle Ricerche, IC-CNR, I-00015 Monterotondo, Roma, Italy

[5] National Research Nuclear University MEPhI (Moscow Engineering Physics Institute), 115409 Moscow, Russia

* Correspondence: A.Bussmann-Holder@fkf.mpg.de

**Abstract:** The basic features of multi-band superconductivity and its implications are derived. In particular, it is shown that enhancements of the superconducting transition temperature take place due to interband interactions. In addition, isotope effects differ substantially from the typical BCS scheme as soon as polaronic coupling effects are present. Special cases of the model are polaronic coupling in one band as realized e.g., in cuprates, coexistence of a flat band and a steep band like in MgB$_2$, crossovers between extreme cases. The advantages of the multiband approach as compared to the single band BCS model are elucidated and its rather frequent realization in actual systems discussed.

## 1. Introduction

Shortly after the upcoming of the BCS theory [1] two independent research teams in the US [2] and Russia [3] introduced important extensions of the theory by considering the possibility that not only one band may contribute to the electron pairing but that two or more bands might be involved. Thereby more realistic band structures can be accounted for than suggested by BCS. In addition, these extensions also allowed for multiple phonon contributions again in accordance with the experimental situation. In spite of these early modifications of the BCS theory it took rather long to find its realization, namely in Nb doped SrTiO$_3$ (STO) [4], where tunneling experiments confirmed the involvement of two electronic bands in superconductivity. In recent studies the interest in this system has erased again, mainly due to the fact that superconducting STO is an extremely low carrier density compound where metallic and superconducting properties appear to be very unlikely.

It took another decade to get back to the issue of multi-band superconductivity which was predominantly advocated by K.A. Müller when considering the pairing mechanism in high temperature superconducting cuprates (HTSC) [5]. His arguments in favor of multiband aspects being realized in HTSC were based on several experimental results which could only be meaningfully understood by postulating that a combination of an s- and a d-wave order coexist in HTSC. Especially, the obvious heterogeneity of this material class has been emphasized by him [6]. Experimentally especially the group around A. Bianconi demonstrated that in the superconducting state of HTSC metallic properties coexist with almost insulating ones in terms of stripes made of alternated distorted and undistorted nanoscale lattice stripes in the $CuO_2$ atomic plane using a fast and local probe, Cu K-edge EXFAS [7] and Cu K-edge resonant elastic X-ray diffraction [8].

With the discovery of $MgB_2$ [9] the multi-band picture achieved a firm basis in the research of superconductivity and was further fostered by the discovery of many more multi-band superconductors. In addition, refined techniques enabled the clear identification of interband and multi-band signatures. While the existence of multi-band superconductors is commonly agreed on, various consequences of it are scarcely addressed. Especially isotope effects vary profoundly from single band theories, and the $T_c$ enhancement due to interband interactions is rarely alluded.

A variety of limiting cases of the multi-band case exist which cover the weak interband coupling limit, the polaronic coupling [10], the coexistence of a steep and a flat band [11], the weak intraband coupling, etc. These are special cases with clear signatures that can be detected experimentally. In addition, Feshbach resonances are possible [12] and the coexistence of a Bose Einstein condensate with BCS type superconductivity [13].

In the following the two-band model is introduced. Polaron formation and its consequences for superconductivity are consecutively discussed. Isotope effects are further addressed and finally the extreme limiting case of the flat band/steep band scenario is considered.

## 2. The Two-Band Model

The Hamiltonian for the two-band model in its simplest form reads:

$$H = H_0 + H_1 + H_2 + H_{12}$$

$$H_0 = \sum_{k_1 \sigma} \xi_{k_1} c^+_{k_1 \sigma} c_{k_1 \sigma} + \sum_{k_2 \sigma} \xi_{k_2} d^+_{k_2 \sigma} d_{k_2 \sigma}$$

where $\xi_{k_i}$ is the band energy $\xi_k = \tilde{\varepsilon} - \varepsilon_k - \mu$, and $\tilde{\varepsilon}_i$ denotes the position of the $c$ and $d$ in bands $i = 1, 2$ and $c^+, c, d^+, d$ are electron creation and annihilation operators in band 1, 2, with spin index $\sigma$. The interband and intraband interactions $H_i$ and $H_{12}$ are explicitly given by:

$$H_1 = -\sum_{k_1 k_1' q} V_1(k_1, k_1') c^+_{k_1+\frac{q}{2}\uparrow} c^+_{-k_1+\frac{q}{2}\downarrow} c_{-k_1'+\frac{q}{2}\downarrow} c_{k_1'+\frac{q}{2}\uparrow}$$

$$H_2 = -\sum_{k_2 k_2' q} V_2(k_2, k_2') d^+_{k_2+q/2\uparrow} d^+_{-k_2+q/2\downarrow} d_{-k_2'+q/2\downarrow} d_{k_2'+q/2\uparrow} \quad (1)$$

$$H_{12} = -\sum_{k_1 k_2 q} V_{12}(k_1, k_2) \{ c^+_{k_1+\frac{q}{2}\uparrow} c^+_{-k_1+\frac{q}{2}\downarrow} d_{-k_2+\frac{q}{2}\downarrow} d_{k_2+\frac{q}{2}\uparrow} + .c. \}$$

The pairing potentials $V_i(k_i, k_i')$ are represented in factorized form as $V_i(k_i, k_i') = V_i \phi_{k_i} \psi_{k_i'}$, where $\phi_{k_i}, \psi_{k_i}$ are cubic harmonics for anisotropic pairing which yields for dimension $d = 2$ and on-site pairing s-wave pairing: $\phi_{k_i} = 1, \psi_{k_i} = 1$, extended s-wave: $\phi_{k_i} = \cos k_x a + \cos k_y b = \gamma_{k_i}$, and d-wave: $\phi_{k_i} = \cos k_x a - \cos k_y b = \eta_{k_i}$ where $a, b$ are the lattice constants along $x$ and $y$ directions, with $a \neq b$ to account for arbitrary lattice symmetry. The individual intraband pairing interactions in the two channels may stem from different sources, namely lattice mediated, magnetic fluctuation driven or some other exotic origin. It is, however, assumed here, that the $c$ channel pairing is of electron-lattice origin, while especially for cuprates the $d$ channel pairing could have a magnetic interaction. With this assumption a combination of s+d wave pairing can result which has an additional $T_c$ enhancement effect as compared to the s+s wave pairing (see below). The important intraband potentials provide the pairwise exchange between the two bands and are dominantly of electron-lattice interaction character. By performing a BCS mean-field analysis of the above equations the gap equations can be derived and the superconducting transition temperature evaluated.

$$H_{red} = \sum_{k_1 \sigma} \xi_{k_1} c^+_{k_1 \sigma} c_{k_1 \sigma} + \sum_{k_2 \sigma} \xi_{k_2} d^+_{k_2 \sigma} d_{k_2 \sigma} + \bar{H}_1 + \bar{H}_2 + \bar{H}_{12}$$

$$\bar{H}_i = -\sum_{k_i'} [\Delta_{k_i'} c^+_{k_i'\uparrow} c^+_{-k_i\downarrow} + \Delta^*_{k_i'} c_{-k_i\downarrow} c_{k_i\uparrow}] + \sum_{k_i, k_i'} V_i(k_i, k_i') < c^+_{k_i\uparrow} c^+_{-k_i\downarrow} >< c_{-k_i'\downarrow} c_{k_i'\uparrow} > \quad (i = 1, 2)$$

(2)

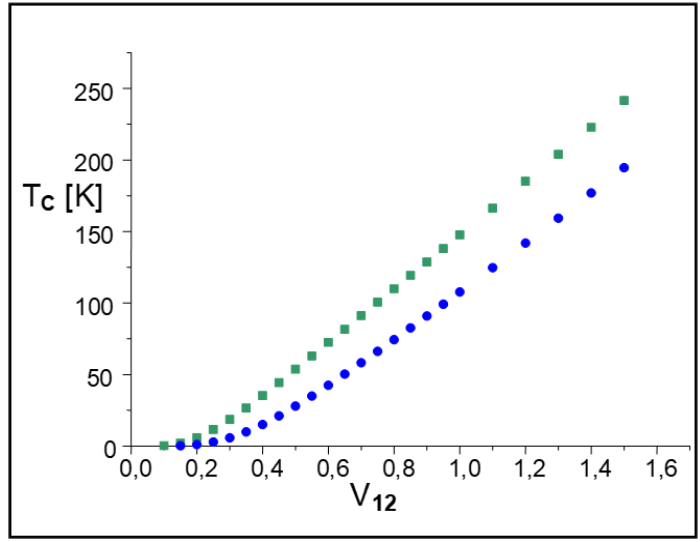

**Figure 1.** Superconducting transition temperature $T_c$ as a function of the interband interaction $V_{12}$. The blue circles refer to s+s, the green squares to s+d pairing symmetry. Note, that the interband interaction ensures a single-valued $T_c$, without it two different values would result—both being much smaller than the one with $V_{12}$ finite. This latter case is, however, physically meaningless.

$$\bar{H}_{12} = -\sum_{k_1,k_2}[V_{12}(k_1,k_2)<c^+_{k_1\uparrow}c^+_{-k_1\downarrow}>d_{-k_2\downarrow}d_{k_2\uparrow} + V_{12}(k_1,k_2)<d_{-k_2\downarrow}>c^+_{k_1\uparrow}c^+_{-k_1\downarrow} +$$
$$V^*_{12}(k_1,k_2)d^+_{k_2\uparrow}d^+_{-k_2\downarrow}<c_{-k_1\downarrow}c_{k_1\uparrow}> + V^*_{12}(k_1,k_2)c_{-k_1\downarrow}c_{k_1\uparrow}<d^+_{k_2\uparrow}d^+_{-k_2\downarrow}> -$$
$$V_{12}(k_1,k_2)<c^+_{k_1\uparrow}c^+_{-k_1\downarrow}><d_{-k_2\downarrow}d_{k_2\uparrow}> - V^*_{12}(k_1,k_2)<c_{-k_1\downarrow}c_{k_1\uparrow}><d^+_{k_2\uparrow}d^+_{-k_2\downarrow}>]$$

and for $i = 2$ $c$ is replaced by $d$. $\xi_{k_i} = \varepsilon_i + \varepsilon_{k_i} - \mu$. Here it is assumed that $<c^+_{k_1+q/2\uparrow}c^+_{-k_1+q/2\downarrow}> = <c^+_{k_1\uparrow}c^+_{-k_1\downarrow}>\delta_{q,0}$ and equivalently for the $d$ operators. In addition the following definitions are introduced:

$$\Delta^*_{k'_i} = \sum_{k_i} V_i(k_i,k'_i)<c^+_{k_i\uparrow}c^+_{-k_i\downarrow}>$$

together with:

$$A^*_{k_1} = \sum_{k_2} V_{12}(k_1,k_2)<d^+_{k_2\uparrow}d^+_{-k_2\downarrow}>, \quad B^*_{k_2} = \sum_{k_2} V_{12}(k_1,k_2)<c^+_{k_2\uparrow}c^+_{-k_2\downarrow}>, \text{ and } V^*_{12} = V_{12}$$

Applying standard techniques we obtain:

$$<c^+_{k_1\uparrow}c^+_{-k_1\downarrow}> = \frac{\bar{\Delta}^*_{k_1}}{2E_{k_1}} \tan\frac{E_{k_1}}{2k_B T} = \bar{\Delta}^*_{k_1}\Phi_{k_1}$$

$$<d^+_{k_2\uparrow}d^+_{-k_2\downarrow}> = \frac{\bar{\Delta}^*_{k_2}}{2E_{k_2}} \tan\frac{E_{k_2}}{2k_B T} = \bar{\Delta}^*_{k_2}\Phi_{k_2}$$

(3)

with $E^2_{k_1} = \xi^2_{k_1} + |\bar{\Delta}_{k_1}|^2, \bar{\Delta}_{k_1} = \Delta_{k_1} + A_{k_1}$ and $E^2_{k_2} = \xi^2_{k_2} + |\bar{\Delta}_{k_2}|^2, \bar{\Delta}_{k_2} = \Delta_{k_2} + B_{k_2}$, which results in the self-consistent set of coupled gap equations:

$$\bar{\Delta}_{k_1} = \sum_{k_1'} V_1(k_1, k_1') \bar{\Delta}_{k_1'} \Phi_{k_1'} + \sum_{k_2} V_{1,2}(k_1, k_2) \bar{\Delta}_{k_2} \Phi_{k_2}$$

$$\bar{\Delta}_{k_2} = \sum_{k_2'} V_2(k_2, k_2') \bar{\Delta}_{k_2'} \Phi_{k_2'} + \sum_{k_1} V_{1,2}(k_1, k_2) \bar{\Delta}_{k_1} \Phi_{k_1} \quad (4)$$

from which the temperature dependencies of the two gaps and the superconducting transition temperature T$_c$ have to be determined. As defined above, $\Phi_{k_i}, i = 1,2$ refers to the pairing symmetry in the respective channel. For cuprates d-wave pairing relates to the planar Cu derived bands, whereas s-wave pairing refers to out-of-plane oxygen bands which are hybridized with the planar bands through interband interactions. The effect of the interband coupling on T$_c$ is demonstrated in Figure 1 for the case of two isotropic s-wave gaps (blue filled circles) and the combination of s+d symmetry (green squares) where a substantial increase in T$_c$ takes place as compared to the s+s case.

## 3. Polaron Effects

Within BCS theory the essential ingredient for the electron pairing mechanism is the electron-phonon interaction. This is typically much smaller than the Fermi energy and is active within a small shell at the top of it. An important observation which led to the discovery of high temperature superconductivity was made by K. A. Müller who primarily was interested in superconductivity in oxides [14]. These have mostly been excluded from the search for new superconducting materials due to their proximity to an insulator. Müller, however, noticed that the few superconducting oxides had a rather small Fermi energy while their T$_c$ was high in comparison to conventional other materials. He could reconcile this finding only by assuming that the electron-lattice interactions must be unusually strong. Such a type of coupling leads to an electron-phonon bound state, namely a polaron. In order to achieve mobility, band like states must be formed. In the crossover region between extremely strong coupling and very weak coupling, the most interesting regime is observed, which is, however, difficult to access. The formation of such a type of quasiparticle has important consequences for the electronic as well as the lattice degrees of freedom. In order to elucidate this in more detail, the typical Hamiltonian for this scenario is given below within the two-band approach discussed above and in relation to HTSC [10]:

$$H = H_d + H_c + H_{cd} + H_L + H_{L-d} + H_{L-c} \quad (5)$$

The first two terms refer to the pure electronic energies in the *c* and *d* channel, the third part corresponds to the interaction between both channels resembling a hybridization term. Further, the non-renormalized lattice part is considered and the two final terms are those where interactions between *c* and *d* electrons (holes) with the lattice are taken into account:

$$H_d = \sum_{i,\sigma} \varepsilon_d d^+_{i,\sigma} d_{i,\sigma} + \sum_{i,j,\sigma,\sigma'} t_{ij}(d^+_{i,\sigma} d_{j,\sigma'} + c.c.) + U \sum_i n_{d,i\uparrow} n_{d,i\downarrow}$$

$$H_c = \sum_{i,\sigma} \varepsilon_c c^+_{i,\sigma} c_{i,\sigma} + \sum_{i,j,\sigma,\sigma'} t_{ij}(c^+_{i,\sigma} c_{j,\sigma'} + c.c.)$$

$$H_{cd} = \sum_{i,j,\sigma,\sigma'} t_{cd}(c^+_{i,\sigma} d_{j,\sigma'} + c.c.)$$

$$H_L = \sum_i \frac{p_i^2}{2M_i} + \frac{1}{2} M \omega^2 Q_i^2 \quad (6)$$

$$H_{L-d} = \sum_{i,j,\sigma,\sigma'} [g n_{d,i} Q_i + \tilde{g}(c^+_{i,\sigma} d_{j,\sigma'} + d^+_{i,\sigma} c_{j,\sigma'}) Q_j]$$

$$H_{L-c} = \sum_{i,j,\sigma,\sigma'} [g n_{c,i} Q_i + \tilde{g}(c^+_{i,\sigma} d_{j,\sigma'} + d^+_{i,\sigma} c_{j,\sigma'}) Q_j]$$

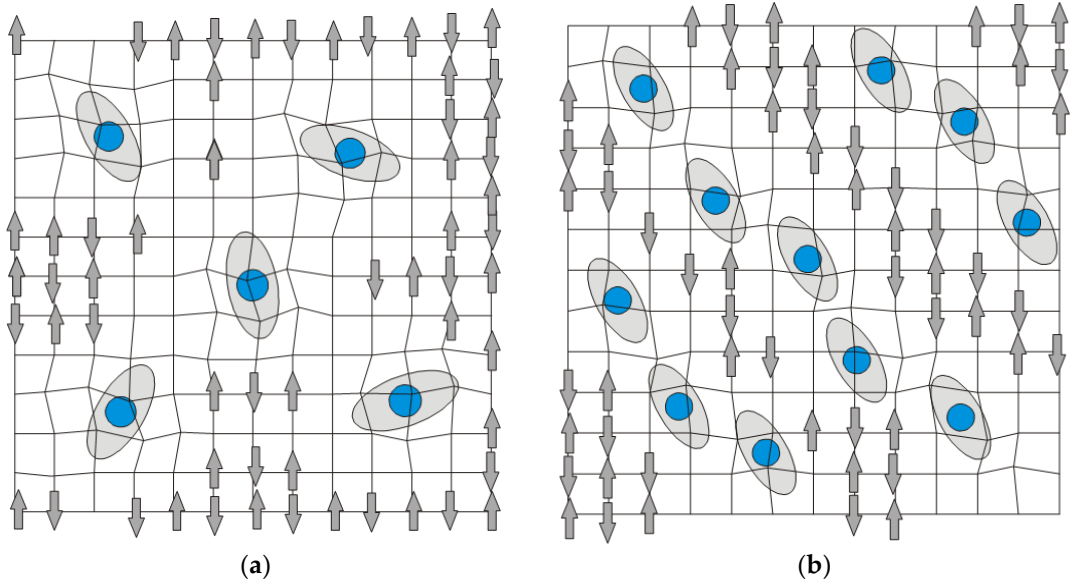

**Figure 2.** Disordered polaron gas (**a**), ordered polaron liquid (**b**) after [18].

Here $\varepsilon$ is the site $i$ dependent electronic energy, $t$ the hopping integral, $U$ the onsite Coulomb repulsion within the antiferromagnetic matrix. This term experiences important renormalizations caused by the electron lattice interaction, being strongly reduced and even be attractive especially within the hole rich distorted regions where metallic droplets are formed which appear as in-gap states in the vicinity of the Fermi surface [15]. Here, $U$ is almost zero with the dominant energy scale being given by the electron-lattice coupling. For simplicity the lattice Hamiltonian is taken to be purely harmonic with $p$ and $Q$ being momentum and conjugate displacement coordinates at site $i$ with frequency $\omega$ and $M$ is the ionic mass. Note however, that higher order anharmonic terms are non-negligible for an appropriate treatment of the lattice dynamics. The coupling between the lattice and the electronic system consists of two terms, namely an onsite coupling proportional to $g$ [16] and an intersite coupling $\tilde{g}$ which provides the interaction between the c and d dominated regions. Basically the two considered regions should be characterized by different coupling constants, but for simplicity these are taken to be the same.

A decoupling of lattice and electronic degrees of freedom can be obtained by a standard Lang-Firsov transformation [17]:

$$\tilde{c}_i = c_i \, exp[\sum_q g_q[b_q^+ - b_q], \quad \tilde{c}_i^+ = c_i^+ \, exp[-\sum_q g_q[b_q^+ - b_q]$$

$$\tilde{d}_i = d_i \, exp[\sum_q g_q[b_q^+ - b_q], \quad \tilde{d}_i^+ = d_i^+ \, exp[-\sum_q g_q[b_q^+ - b_q] \quad (7)$$

$$\tilde{b}_q = b_q + \sum_q g_q d_i^+ d_i, \quad \tilde{b}_q^+ = b_q^+ + \sum_q g_q d_i^+ d_i$$

and a corresponding transformation for the *c*-channel. As a consequence of this transformation a rigid band shift appears in the site energies, i.e., $\varepsilon \to \tilde{\varepsilon} = \varepsilon - \Delta^*$, with $\Delta^* = \frac{1}{2N} \sum_q \frac{g_q^2}{\hbar \omega_q}$, the hopping integrals are exponentially reduced as $t \to \tilde{t} = t \, exp[-g^2 \, coth \frac{\hbar \omega}{2kT}]$. In addition, lattice mediated $n_c n_d$ density-density interactions between $c$ and $d$ bands are a consequence, which facilitate multiband superconductivity. The new quasiparticles are now dressed and carry a local distortion thereby reducing reduces their mobility. This means that the band states related to the doped holes within the metallic droplets lose dispersion, whereas those related to the *d* holes, which are nearly localized in the undoped case, gain dispersion due to their interaction with the *c* states. A strange metal phase is the consequence.

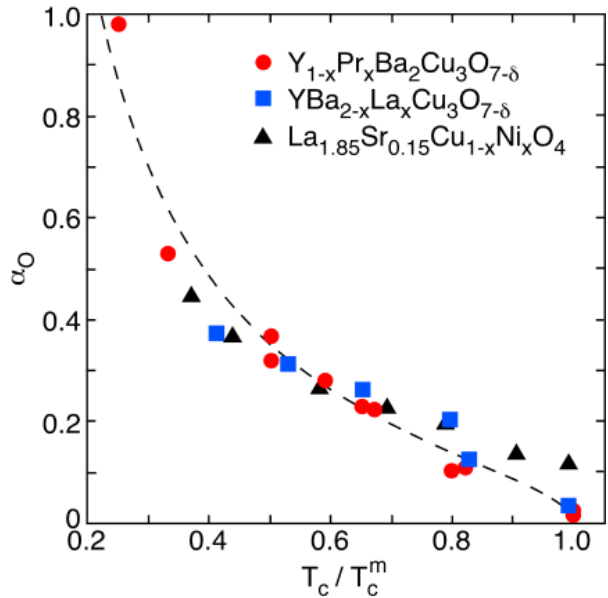

**Figure 3.** Oxygen isotope-effect exponent $\alpha_O$ versus $T_c/T_c^m$ for various families of cuprate HTS ($T_c^m$ denotes the maximum $T_c$ for a particular family). The dashed line is a guide to the eye and indicates how $\alpha_O$ increases with reduced $T_c/T_c^m$ after [18].

At high temperature the polarons are randomly distributed over the lattice, forming a liquid (Figure 2) [18]. A transition to an ordered state takes place with decreasing temperature and ordering into patterns (stripe segments) becomes energetically favorable in order to minimize the interaction energy attributed to strain fields.

It is, however, important to mention that the patterning remains dynamic in order to ensure the interplay between *c* and *d* dominated regions. Static stripes have been observed at around 1/8 doping in LaSrCuO₄ where superconductivity is suppressed [19]. The transition from the polaron "liquid" to the polaron "glass" takes place at a characteristic temperature which we identify as the pseudogap temperature $T^*$ which corresponds to a transition from a disordered phase to an ordered one. Here a coupling of the disordered polaron state (liquid) to the strain fields takes place, and the transition temperature becomes a function of the coupling $g$ of the polarons through the strain fields, the number of polarons, i.e., the doping level $\delta$, and the energy $\varepsilon_p$ to remove a single polaron from its ordered arrangement, i.e., $k_B T^* = (\varepsilon_p - g/2)/\ln\delta$. An implicit equation for $T^*$ can be deduced from the renormalization of the lattice degrees of freedom (Equation (7)) which results in renormalized local modes with frequency $\widetilde{\omega}_q^2 = \omega_0^2(q) - g^2 \sum \frac{1}{\varepsilon(k)} tanh\frac{\varepsilon(k)}{kT}$ which soften upon approaching $T^*$ where $\omega_0^2(q)$ is the non-renormalized normal mode frequency. The corresponding temperature is then given by: $\frac{\omega_0^2}{g^2} = \int \frac{dk}{\varepsilon(k)} tanh\frac{\varepsilon(k)}{kT^*}$. From both expressions for $T^*$ an isotope effect can be derived which is sign reversed and large.

## 4. Isotope Effects

The verification of the BCS theory was definitely supported by the observation of an isotope effect on $T_c$. Since then one of the first experiments to identify the electron (hole) pairing mechanisms in newly discovered superconductors, was the measurement of an isotope effect. Even though many conventional superconductors show deviations from the BCS prediction of $-d\ln T_c/d\ln M = \alpha = 0.5$, where $M$ is the ionic mass, it was mostly accepted that these are due to e.g., increased Coulomb interactions, density of states effects, etc. In HTSC an apparent discrepancy between the BCS value and the actually measured one was reported early on [20–22], and, in addition, a strong dependence of $\alpha$ on doping observed. Basically, $\alpha$ is almost zero at optimum doping and increases rapidly with decreasing doping to even exceed 1 in the underdoped regime [23]. Such a behavior is clearly not understandable within any simple phonon mediated pairing mechanism, is, however, a clear signature for a lattice involvement. Besides of cuprate HTSC, also pnictides and $MgB_2$ show unconventional isotope effects and are shortly mentioned below.

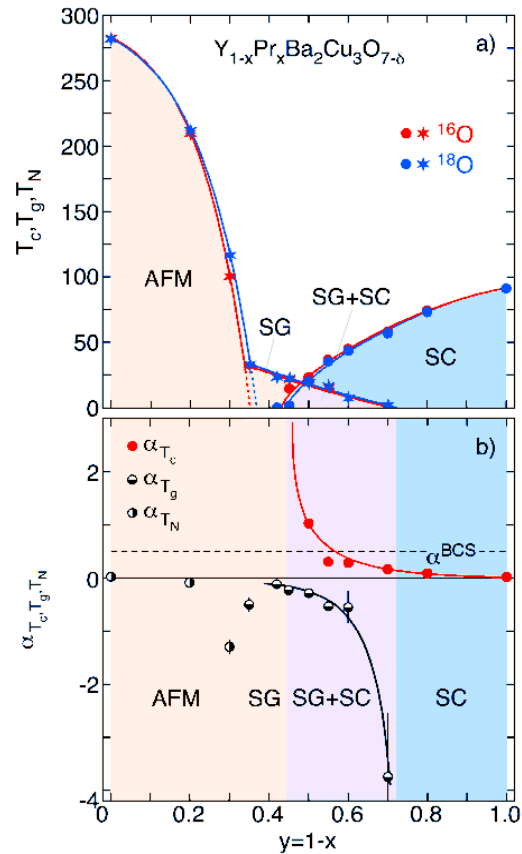

**Figure 4.** (**a**) Dependence of the superconducting transition temperature $T_c$, the spin glass transtion temperature $T_g$, and the Néel Temperature $T_N$ for $^{16}O/^{18}O$ substituted $Y_{1-x}Pr_xBa_2Cu_3O_{7-\delta}$ on the Pr content $y = 1 - x$. The solid lines are guides to the eye. The areas denoted by AFM, SG, and SC represent the antiferromagnetic, the spin glass, and the superconducting regions, respectively. In the region SG + SC spin-glass magnetism and superconductivity coexist. (**b**) Dependence of the corresponding oxygen isotope exponent (OIE) exponents $\alpha(T_c)$, $\alpha(T_g)$, and $\alpha(T_N)$ as a function of $y = 1 - x$. The dashed line corresponds to the BCS value $\alpha(T_{c\text{-BCS}}) = 0.5$. The solid lines are guides to the eye. After [35].

*4.1. Cuprates*

Experiments on the isotope effect in cuprates have been performed early on, and have—in view of the absence of it at optimum doping—not been taken too seriously in identifying a pairing mechanism. Nevertheless, it was measured for diverse cuprates rather systematically as a function of doping in (Figure 3).

Besides of measuring the "global" oxygen isotope effect, it was also possible to determine it as a function of the local oxygen ion position, the so-called site selective oxygen isotope effect [24,25]. In view of the fact that the apical oxygen ions were assumed to play an important role for the pairing mechanism, such measurements are a key to identifying their role. Interestingly, the largest isotope effect stems from the in-plane oxygen ions whereas the apical ions almost do not contribute to it. Besides of this unique property, another unexpected within BCS theory isotope effect on the magnetic field penetration depth has been detected [26]. Since the penetration depth is directly related to the superfluid carrier density and their inverse effective mass, this effect suggests that polaron formation is causing it [27]. A more consistent explanation has been offered in a variety of work where polaronic band narrowing has been shown to be responsible for almost all observed isotope effects [10,18].

Further isotope effects have subsequently been explored where a huge and inverse one on the the pseudogap temperature $T^*$ [28–31] has been discovered. Opposite to expectations from classical lattice dynamics, $T^*$ shifts to higher temperatures with the heavier isotope. Again, polaron formation can explain this finding, namely, in terms of local mode phonon softening caused by polaron renormalization effects [32–34].

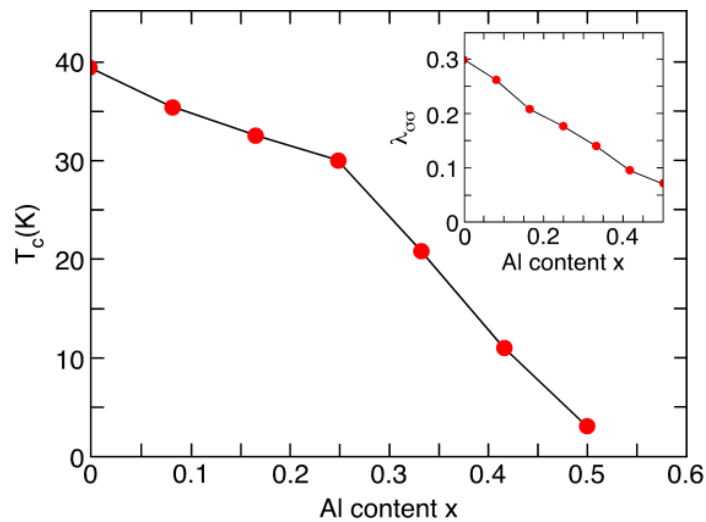

**Figure 5.** Superconducting transition temperature $T_c$ as a function of Al doping x in $Mg_{1-x}Al_xB_2$. The full circles are experimental data [40], the line is the calculated x-dependence of $T_c$. The inset to the Figure shows the calculated x-dependent intraband coupling $\lambda_{\sigma\sigma}$ related to the σ-band.

Finally, the phase diagram of cuprates has been investigated for further isotope effects and the results are summarized in [35] in Figure 4 where all phases of the phase diagram are displayed. In the undoped antiferromagnetic phase the isotope effect on $T_N$ increases from almost zero to large negative values and continues to remain negative upon entering the spin glass phase. In the coexistence regime of spin glass phase and superconductivity an isotope effect consists of two-components suggesting phase separation: spin glass freezing temperature $T_{SG}$ shows an inverse isotope effect whereas a conventional is observed for $T_c$.

*4.2. MgB$_2$*

Superconductivity in magnesium diboride [9] was early on identified to stem from two bands, namely a π and a σ-band [36–45]. Upon doping $MgB_2$ with Al or Sc superconductivity persists over a rather broad range. The phase diagram of $Mg_{1-x}Al_xB_2$ is shown in Figure 5. This system then admits to explore in detail the tuning of the chemical potential by strain and charge transfer near an electronic topological Lifshitz transition for the appearing of small tubular spots of the σ Fermi surface [36,42] which are revealed by the doping evolution of the superconducting energy gaps [38,42].

Two phonon modes play a significant role to the electron pairing in $MgB_2$, namely the $E_{2g}$ mode energy in the σ channel and an averaged phonon frequency $\omega_{ln}$ in the π channel [43]. The BCS isotope exponent $\alpha = 0.5$ is only observed if both phonon frequencies depend simultaneously on the same ionic mass. This is, however, not the case in Al doped MgB$_2$, and within a two-band model a reduced isotope exponent is expected as also observed experimentally [44].

A theoretical derivation of the isotope effect is based on the assumption that an average frequency $\omega_{ln}$ and the experimental $E_{2g}$ phonon mode frequency are dominant. The latter mode hardens with Al doping and splits at low doping levels into a hard and a soft component [43]. In order to avoid additional complications in the theoretical modelling the splitting of this mode is ignored and an interpolation scheme between the soft and the hard mode regimes used.

Two cases are considered for the calculation of the isotope effect on $T_c$, namely the harmonic isotope shift of the $E_{2g}$ phonon mode with $\omega_{E_{2g}} \approx M_B^{-0.5}$ and the reduced one where $\omega_{E_{2g}} \approx M_B^{-0.4}$. The results of the calculation are shown in Figure 6 together with the isotope effect on the $E_{2g}$ phonon mode [44].

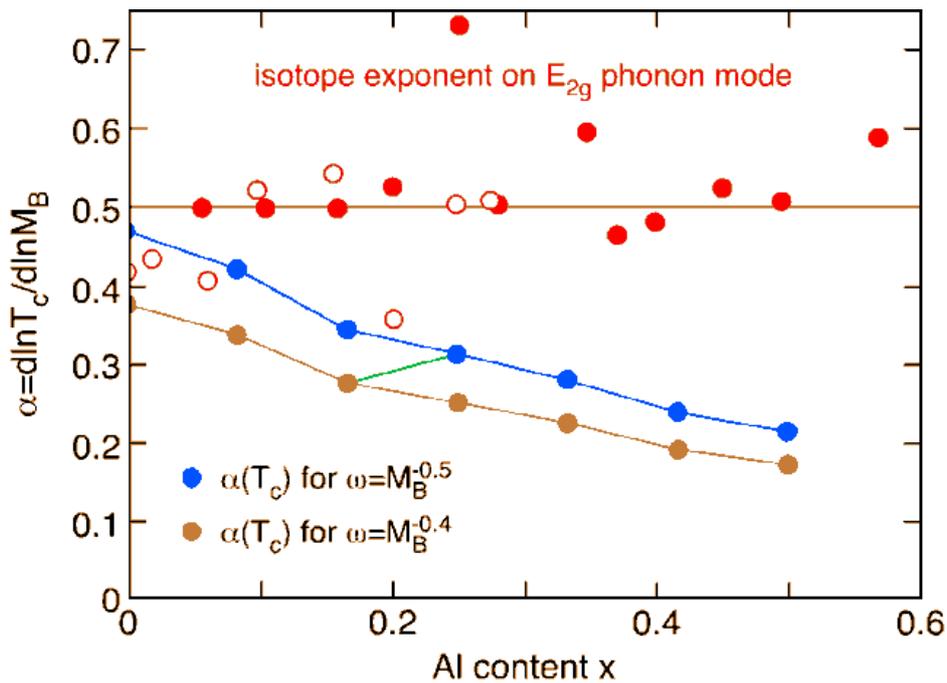

**Figure 6.** Isotope exponent $\alpha$ on $T_c$ as a function of Al doping x in Mg$_{1-x}$Al$_x$B$_2$. The blue circles refer to $\alpha$ as obtained from a harmonic mass dependence of the $E_{2g}$ mode frequency, the brown circles are the same, however assuming that the $E_{2g}$ mode frequency is $\approx$ M$_B^{-0.4}$. Between x = 0.15 and 0.25 a crossover between both behaviors is expected as indicated by the green line. The full and open red circles are the experimentally obtained isotope effects on the $E_{2g}$ mode frequency [40]. For details refer to Ref. [43].

Further searches for more exotic isotope effects in MgB$_2$, as, e.g., on the penetration depth [44-45], which would point to a polaronic pairing glue, were unsuccessful. This lead to the conclusion that MgB$_2$ is a rather conventional phonon mediated superconductor where the π-electron related band plays a minor role and the coupling to the $E_{2g}$ phonon mode in the σ-band is the driving mechanism. In spite of this conclusion the role of the π-band cannot be neglected since is essential in enhancing $T_c$ by involving the electrons for the pairwise exchange between the two bands and the crucial increase in $T_c$.

### 4.3. Fe Based Superconductors

In 2008 a new family of multiband and multi-gap high temperature superconductors called iron based superconductors [46]. Quite opposite to cuprates where oxygen isotope exchange can be performed rather directly, an isotope replacement in Fe based superconductors [47-51] is very demanding, since situ exchange of Fe or other constitutes such as As by their isotopes is impossible. A way out of this dilemma is to perform statistics over many samples with the same preparation techniques and growth conditions [49].

Two groups have independently and almost simultaneously performed the Fe isotope exchange in nominally identical FeAs based compounds [50,51]. However, the outcome of these experiments was highly controversial since a conventional isotope exponent of $\alpha \approx 0.3$ was reported in [50] and a sign reversed one of $\alpha \approx -0.3$ [51]. Shortly afterwards the former group carried out another isotope experiment with similar results as the first but with $\alpha$ being slightly reduced [52]. In view of this rather unusual situation an independent third study was undertaken, by using statistics and a conventional isotope effect found in accordance with the first publication (Figure 7). This last experiment demonstrated in addition that the isotope exchange was accompanied by small structural modifications, which—in view of the sensitivity of $T_c$ to tiny structural alterations—could substantially influence the isotope effect. Consequently, these structural changes have been investigated in deeper detail [53] with the result, that if these are properly taken into account the isotope effect reduces to the same value as observed in the first study, namely, $\alpha \approx 0.3$ (Figure 8).

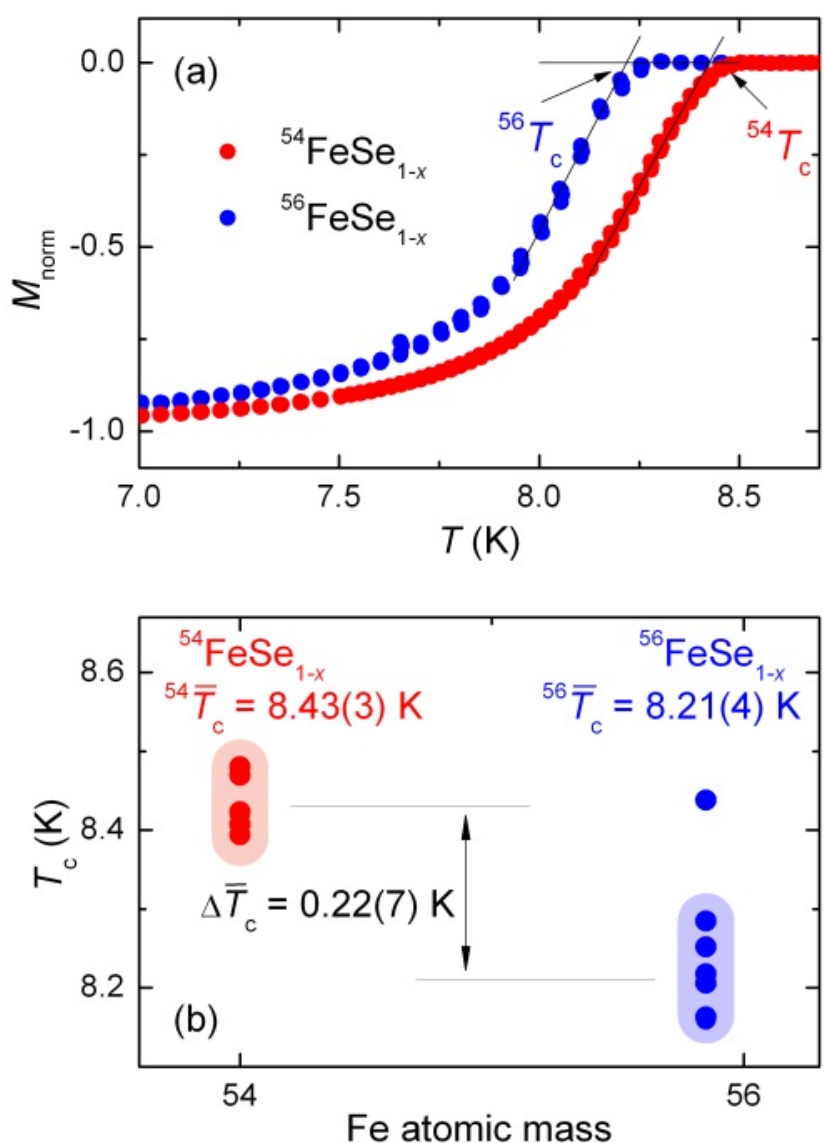

**Figure 7. a)** The superconducting transition temperature $T_c$ as a function of the Fe atomic mass for a series of $^{54}FeSe_{1-x}$ and $^{56}FeSe_{1-x}$ samples studied in Ref. [49]. b) The $T_c$'s fall into two regions marked by the colored stripes, indicating an $^{54}Fe/^{56}Fe$ isotope shift of $\Delta T_c = 0.85(15)$ K. After Ref. [49].

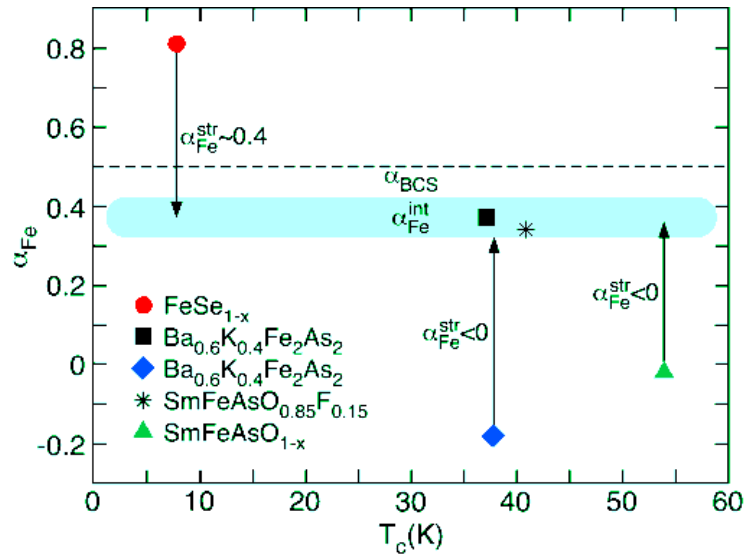

**Figure 8.** Fe isotope effect exponent α(Fe) as a function of $T_c$ for the iron-based superconductors discussed in this work: $FeSe_{1-x}$, $Ba_{0.6}K_{0.4}Fe_2As_2$ and $SmFeAsO_{0.85}F_{0.15}$, $Ba_{0.6}K_{0.4}Fe_2As_2$], and $SmFeAsO_{1-x}$ [53]. Arrows indicate the correction caused by structural effects. Note that α($Fe_{int}$) ≈ 0.35–0.40 for all samples (α$_{BCS}$ = 0.5 denotes the weak-coupling BCS value). After [54].

## 5. The Steep Band/Flat Band Model

A special case of the Hamiltonian Equation (1) is realized by the steep band/flat band scenario [11-16,55–61]. This corresponds to a band with itinerant Fermi liquid type character and a localized dispersion-less polaronic one [5–8,30,62–67] which admits to approximate the band energies like: $\varepsilon_1 = const = B$; $\varepsilon_2(k) = k^2/2m$ . A large but finite density of states $N_i$ ($i$ = 1, 2) is attributed to both bands and defining $\sqrt{N_1/N_2} = C$ , the $T_c$ defining equation can be formulated like:

$$1 = \lambda_{11} \int_0^{\hbar\omega_1} \frac{1}{B} tanh\left[\frac{B}{2kT_c}\right] d\varepsilon_1 + \lambda_{12} \frac{1}{C} \int_0^{\hbar\omega_{1,2}} \frac{1}{\frac{k^2}{2m}} tanh\left[\frac{\frac{k^2}{2m}}{2kT_c}\right] dk$$
$$1 = \lambda_{22} \int_0^{\hbar\omega_2} \frac{1}{k^2/2m} tanh\left[\frac{k^2/2m}{2kT_c}\right] dk + \lambda_{21} C \int_0^{\hbar\omega_{1,2}} \frac{1}{B} tanh\left[\frac{B}{2kT_c}\right] d\varepsilon_1 \quad (8)$$

with $\lambda_{ii}$ and $\lambda_{ij}$ being the product of the pairing potential and the density of states.

The integrals are replaced by the logarithmic BCS expression and with $\lambda_{12} = \lambda_{21}$ can be solved like:

$$1 = \lambda_{11} \int_0^{\hbar\omega_1} \frac{1}{B} tanh\left[\frac{B}{2kT_c}\right] d\varepsilon_1 + \lambda_{12} \frac{1}{C} \int_0^{\hbar\omega_{1,2}} \frac{1}{\varepsilon_2(k)} tanh\left[\frac{\varepsilon_2(k)}{2kT_c}\right] d\varepsilon_2$$
$$1 = \lambda_{22} \int_0^{\hbar\omega_2} \frac{1}{\varepsilon_2(k)} tanh\left[\frac{\varepsilon_2(k)}{2kT_c}\right] d\varepsilon_2 + \lambda_{12} C \int_0^{\hbar\omega_{1,2}} \frac{1}{B} tanh\left[\frac{B}{2kT_c}\right] d\varepsilon_1 \quad (9)$$

$$1 = \frac{\lambda_{11}\hbar\omega_1}{B} tanh\left[\frac{B}{2kT_c}\right] + \lambda_{12}\frac{1}{C} ln\left[\frac{1.13\hbar\omega_{1,2}}{kT_c}\right]$$
$$1 = \lambda_{22} ln\left[\frac{1.13\hbar\omega_2}{kT_c}\right] + \frac{\lambda_{12} C \hbar\omega_{1,2}}{B} tanh\left[\frac{B}{2kT_c}\right] \quad (10)$$

The two approximate solutions are 1. $x \ll 1$: $tanh(x) = x$ , 2. $x \gg 1$: $tanh(x) = 1$ .

$$1 = \lambda_{11} \frac{\hbar\omega_1}{2kT_c} + \lambda_{12}\frac{1}{C} ln\left[\frac{1.13\hbar\omega_{1,2}}{kT_c}\right] = \lambda_{12} C \frac{\hbar\omega_{1,2}}{2kT_c} + \lambda_{22} ln\left[\frac{1.13\hbar\omega_2}{kT_c}\right] \quad (11)$$

yielding an implicit relation for $T_c$.

$$1 = \frac{\lambda_{11}\hbar\omega_1}{B} + \lambda_{12}\frac{1}{C}ln\left[\frac{1.13\hbar\omega_{1,2}}{kT_c}\right] = \lambda_{12}C\frac{\hbar\omega_{1,2}}{B} + \lambda_{22}ln\left[\frac{1.13\hbar\omega_2}{kT_c}\right] \quad (12)$$

which can be solved explicitly:

$$kT_c = [1.13\hbar\omega_2 exp\left(\frac{-\lambda_{22}}{\lambda_{12}\frac{1}{C} - \lambda_{22}}\right)$$
$$+ 1.13\hbar\omega_{1,2} exp\left(\frac{-\lambda_{12}}{\lambda_{22} - \lambda_{12}C}\right)]exp\left(\frac{\lambda_{12}\frac{1}{C}\hbar\omega_{12} - \lambda_{11}\hbar\omega_1}{B(\lambda_{22} - \lambda_{12}C)}\right) \quad (13)$$

Equation (13) is distinctly different from the one obtained by by Suhl, Matthias and Walker (SMW) [51] where the phonon cutoff energies are identical for the different pairing channels, and the band dispersion remains undefined.

Two possibilities are considered in the following, namely, that the flat band is realized either in the strong coupling band or in the weak coupling band. For both cases the interband interaction plays the dominant role in enhancing $T_c$. However, also the band 1 related intraband coupling is decisive and the corresponding $T_c$ as a function of $\lambda_{11}$ displayed in Figure 9.

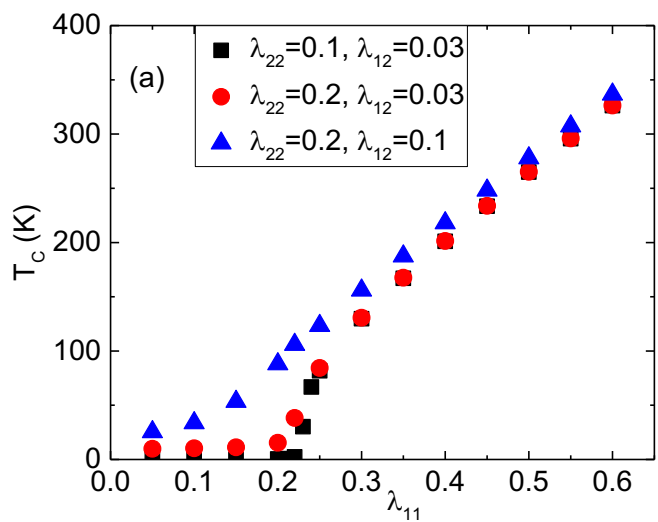

**Figure 9.** $T_C$ as a function of the intraband coupling $\lambda_{11}$ which is related to the strong coupling flat polaronic band. The black squares refer to the parameters $\lambda_{22} = 0.1, \lambda_{12} = 0.03$, red circles: $\lambda_{22} = 0.2, \lambda_{12} = 0.03$, blue triangles: $\lambda_{22} = 0.2, \lambda_{12} = 0.1$.

The amazing and generic dependence of $T_c$ on $\lambda_{11}$ is shown in Figure 9. $T_c$ remains negligibly small up to values of $\lambda_{11} = 0.2$ with the exception of the increased interband coupling $\lambda_{12} = 0.1$ which induces also for small values of $\lambda_{11}$ considerable increases in $T_c$. This in contrast to $\lambda_{22}$ which—when doubled—does not affect much the $T_c$ versus $\lambda_{11}$ dependence. For all values $\lambda_{11} > 0.2$ a rapid increase in $T_c$ takes place to reach values of more than 300 K for $\lambda_{11} = 0.6$. This observation emphasizes the importance of the polaronic band and demonstrates that a leading role in $T_c$ enhancements relates to the intraband polaronic coupling. The development of $T_c$ with $\lambda_{11}$ is distinctly different from the case shown in Figure 1 where a soft parabolic enhancement of it with increasing interband coupling takes place.

## 6. Conclusions

In the above chapters we have shown that extensions of the BCS theory to more than one band have strong influences on $T_c$, the isotope effects, the magnitude of the gap to $T_c$ ratio. In any case an enormous enhancement of $T_C$ can take place which is beyond any conventional strong or weak coupling case. This enhancement effect is suggested to occur also in the extremely low carrier density system STO where neither metallicity nor superconductivity should occur. However, upon treating this compound as a realization of the steep band/flat band scenario offers a way out of this dilemma. By including polaronic effects especially the isotope effect is affected which deviated strongly from the BCS value. Besides of varying from almost zero to exceeding large the BCS limit, additional isotope effects on the pseudogap temperature $T^*$, the penetration depth and the whole phase diagram of cuprates are a consequence. This is in contrast to Fe based superconductors where almost conventional phonon type interactions occur also if bond fluctuations in a double well have been observed [68,69] and in doped diborides phonon softening [39] thermal conductivity [70] and anomalous temperature dependent lattice effects [71] show the presence of a Kohn anomaly. These results show that cuprates, diborides [72–74] and iron based superconductors [75] are close to Lifshitz transitions for the appearance of a Fermi surface spot. The special case of the steep band/flat band scenario in the multiband superconductors admits for further increases in $T_C$ and it has been proposed to be applicable to hydrogen based superconductors [76–79] where pressure tunes the system at electronic topological transitions [80,81], with the formation of a complex scenario made of a nanoscale phase separation where free particles and polarons coexist in a scale invariant texture [82].

**Author Contributions:** A.B.H. has done the calculations and editing; H.K., A.S., and A.B. have contributed to the editing. All authors equally contributed to this work.

**Funding:** This research was funded by superstripes-onlus and Max-Planck-Institute.

**Conflicts of Interest**: The authors declare no conflicts of interest.